\documentclass[12pt,preprint]{aastex}

\slugcomment{Submitted to The Astrophysical Journal}

\shorttitle{Magnetically Arrested Disks}
\shortauthors{I. V. Igumenshchev}

\begin{document}









\title{MAGNETIC INVERSION AS A MECHANISM FOR
THE SPECTRAL TRANSITION OF BLACK HOLE BINARIES}

\author{Igor V. Igumenshchev}
\affil{Laboratory for Laser Energetics, University of Rochester\\
250 East River Road, Rochester, NY 14623}
\email{iigu@lle.rochester.edu}

\begin{abstract} 

A mechanism for the transition between
low/hard, high/soft, and steep power law (SPL)
spectral states in black hole X-ray binaries
is proposed. The low/hard state is explained by the development of
a magnetically arrested accretion disk attributable to the accumulation of a vertical
magnetic field in a central bundle. This disk forms powerful
jets and consists of thin spiral accretion streams of a dense
optically thick plasma surrounded by hot, magnetized, optically thin corona,
which emits most of the energy in hard X-rays.
State transition occurs because of the quasi-periodic or 
random inversion of poloidal magnetic fields
in the accretion flow supplied by the secondary star.
The inward advection of the inverted field results in a temporal
disappearance of the central bundle
caused by the annihilation of the opposed fields and restoration of
the optically thick disk in the innermost region. 
This disk represents the high/soft state.
The SPL state develops at the period of intensive field annihilation
and precedes the high/soft state.
The continuous supply of the inverted field leads to
a new low/hard state because of the formation of another
magnetically arrested disk.

\end{abstract}


\section{Introduction}

Accretion disks orbiting black holes (BHs) in X-ray binary systems
demonstrate a complicated time-dependent behavior, which is
typically described as the quasi-periodic transition between
three emission states: 
the low/hard, high/soft, and steep power law (SPL, or otherwise, very high) states 
(Remillard \& McClintok 2006, hereafter RM06; Done, Gierli\'nski, \& Kubota 2007).
The low/hard state is characterized by a strong 
hard X-ray emission component in the 2-20 keV range, which takes
$\ga 80\%$ of the total flux.
A weak, but sizable, thermal component from a cold dense plasma
and quasi-periodic oscillations (QPOs) in X-rays may be present or absent.
This state is associated with the formation of quasi-steady radio jets
(Gallo, Fender, \& Pooley 2003).
The high/soft state shows the dominant thermal component,
which is generally consistent with theoretical predictions from
the ``standard" optically thick accretion disk model (Shakura \& Sunyaev 1973).
A hard X-ray component is usually present in this state, 
but is limited to $< 25\%$ of the total flux.
The SPL state shares some common properties with the high/soft state, such as
the thermal component. However, the SPL state is clearly
distinguished by the strength of its power law component and association with
high-frequency QPOs (RM06).

While the high/soft state is reasonably well understood theoretically
(e.g., Kubota {\it et al.} 2005), 
the nature of the low/hard and SPL states is still a matter of debate.
Basically, two classes of phenomenological models
were proposed to explain the low/hard state:
``truncated-disk" and ``corona-and-disk" models.
The truncated-disk model assumes that an optically thick disk is truncated
at some inner radius $R_{\rm tr}$ and the central region, $R<R_{\rm tr}$, is filled
with a hot (about the virial temperature), optically thin plasma,
which produces hard X-rays (e.g., Zdziarski \& Gierli\'nski 2004).
Some studies postulate advection-dominated accretion 
flows (ADAFs, see Narayan \& Yi 1994; Abramowicz {\it et al.} 1995) 
or other types of hot accretion disk solutions
(Shapiro, Lightman, \& Eardley 1976; Blandford \& Begelman 1999;
Narayan, Igumenshchev, \& Abramowicz 2000), 
as sources of this hot plasma
(Poutanen, Krolik \& Ryde 1997; Esin, McClintock, \& Narayan 1997;
Esin {\it et al.} 1998, 2001; 
D'Angelo {\it et al.} 2008). 
Although such studies can provide excellent fits to the combined optical,
UV, and X-ray data, 
the nature of accretion flows in the central region of luminous
sources (with $L\ga 10^{-3}L_{\rm Edd}$, where $L_{\rm Edd}$
is the Eddington luminosity) is unexplained.
Mechanisms that produce truncated disks have been discussed by
Honma (1996) and Manmoto \& Kato (2000), who assumed a radial conductive 
energy transport that leads to an evaporation of thin disks, 
and by Meyer, Liu, \& Meyer-Hofmeister (2000), Spruit \& Deufel (2002), and
Dullemond \& Spruit (2005), who considered a vertical evaporation process.

In the corona-and-disk models, the optically thick disk remains untruncated
and the hard X-rays are generated in a hot, patchy disk corona because of inverse
Compton scattering of soft photons that come from the underlying disk
(Liang \& Price 1977; Galeev, Rosner, \& Vaiana 1979; 
Haardt, Maraschi, \& Ghisellini 1994).
To reasonably describe observations, the hot corona should dissipate
(probably in magnetic flares) a significant fraction ($\ga 50\%$)
of the binding energy of the accretion mass.
It is assumed that this energy is transported from disk
to corona by a magnetic field, but the exact mechanism is unknown
(Merloni \& Fabian 2001; Uzdensky \& Goodman 2008).

Both truncated-disk and corona-and-disk models have advantages and
disadvantages to explain the observed data (see Done {\it et al.} 2007). 
To overcome the disadvantages, several more
sophisticated models were developed, some of which included the elements of both
classes of models described above (Taam {\it et al.} 2008) and
others 
considered
jets that emit in radio and X-rays 
(Markoff, Falke, \& Fender 2001; Ferreira {\it et al.} 2006).
The transition between emission states 
in the truncated-disk models
is assumed by means of shifting the truncation radius
$R_{\rm tr}$ to or out of the BH
(e.g., Petrucci {\it et al.} 2006).
The parameter, or a set of parameters, that
triggers the transition is unknown. 
For example, the triggering
mechanism's dependence on the luminosity, or accretion rate, is ruled out
by observations
that show that many sources can have about same total luminosity
at different states (Done \& Gierli\'nski 2003).

This letter proposes a detailed mechanism for spectral
variability of BH X-ray binaries. The mechanism uses
a recently found dynamical model of magnetized accretion disks
(Igumenshchev 2008, hereafter I08) to explain the low/hard
and SPL states,
and assumes that an inversion of vertical magnetic fields in the disks
triggers the state transition.
Similar magnetic inversion mechanisms were considered in recent studies
of BH X-ray transients and active galactic nuclei (Livio, Pringle, \& King 2003; 
Tagger {\it et al.} 2004) and discussed by 
Igumenshchev, Narayan, \& Abramowicz (2003, hereafter INA03) 
in a more general context.
The present study specifically focuses on the structure and dynamics of accretion disks 
that experience magnetic inversion.
Details of the model of the low/hard state and its possible observational
implications are discussed in Sec.~II. 
Section~III describes the transition mechanism,
presents supporting simulation results, and briefly discusses the nature
of the SPL state.
The letter is summarized in Sec.~IV.

\section{Model for the Low/Hard State}

The key ingredient of the model is a bipolar magnetic field that
is supplied with accretion flows at the outer boundary.
Such flows were modeled by I08 using combined 2D/3D
numerical MHD simulations.
The simulations assume a permanent injection of mass and 
poloidal magnetic field
into a slender equatorial torus located at $R_{\rm inj}$ near the outer boundary 
$R_{\rm out}=220\,R_{g}$, where $R_{g}=2GM/c^2$
and $M$ is the BH mass.
The injected mass and field form a magnetized accretion disk that
carries the vertical magnetic field inward.
At the inner numerical boundary,
$R_{\rm in}=2\,R_{\rm g}$, the mass is absorbed by the BH,
but the vertical field, which cannot be absorbed,
is accumulated in a central bundle. 
When the field $B_0$ in the bundle approaches the equipartition level,
\begin{equation}
{B_0^2\over 8\pi} \sim \rho{GM\over R_{\rm g}},
\end{equation}
where $\rho$ is the density,
the accretion flow is arrested by the field.
Further accumulation of the field
results in the growth of the outer radius
of the arrested region, $R_{\rm m}$.
This type of accretion flow
was anticipated in previous theoretical and numerical studies
(Bisnovatyi-Kogan \& Ruzmaikin 1974, 1976; INA03; Narayan, Igumenshchev, \& Abramowicz 2003)
and was called the magnetically arrested disk.
Most of the volume in such a disk is filled with a hot
(with temperature $T\sim GMm_{\rm p}/R$, where $m_{\rm p}$ is the proton mass), 
highly magnetized
(with plasma $\beta\ll 1$), low-density plasma.
The accretion disk interacts with this plasma at $R_{\rm m}$
and forms geometrically thin and dense streams, which
accrete into the BH on spiral trajectories,
flowing around  low-density ``magnetic islands" (see Figs.~4 and 6 in I08).
The flow pattern 
is highly variable because of the development
of Rayleigh--Taylor and Kelvin--Helmholtz instabilities.
The infall velocity of the dense plasma in streams is a large
fraction ($\sim 0.5$) of the free-fall velocity,
which is why the streams are geometrically thin and 
occupy only a small fraction of the volume in a magnetically arrested disk.
The high infall velocity also indicates that there is an efficient
``braking" of the rotating flow by means of the vertical field. 
This braking results in the
transfer of the rotational energy of the flow to the field energy and the release
of the latter energy in the form of bipolar Poynting jets.
The simulations of I08 demonstrated
that $\sim 1\%$ of $\dot{M}c^2$, where $\dot{M}$ is the
mass accretion rate, can go into the jets
in the case of magnetically arrested disks orbiting non-rotating BHs.
Note that the jet power can be substantially enhanced in the case of
fast rotating BHs with properly aligned spins
because of the ``ergospheric disk" mechanism
(Punsly \& Coroniti 1990; Punsly, Hirose, \& Igumenshchev 2009).

The magnetically arrested disk model is a promising candidate
to explain BH binaries in the low/hard state
because of its specific properties.
This model can have an outer optically thick disk that is 
naturally truncated at $R_{\rm tr}= R_{\rm m}$,
similar to what is postulated in the phenomenological truncated-disk model.
Inside $R_{\rm m}$, a soft X-ray emission from the disk is suppressed because of
fast accretion of the mass in spiral streams,
which an optically thick medium cannot efficiently radiate because of
the long Thomson-scattering diffusion time of photons 
in comparison with the accretion time (I08).
Instead, the binding energy of the accretion
flow is transformed into the energy of torsional MHD waves that
feed the Poynting jets and
heat the low-density plasma, which surrounds the dense accretion streams,
via magnetic flares. 
This plasma can emit hard X-rays via bremsstrahlung and Compton scattering
of soft photons from the dense streams.
Such emission properties plus the presence of the
dense plasma in the vicinity of the BH cause the magnetically arrested disk model to resemble
to another phenomenological model for the low/hard state, corona-and-disk model.

The existence of magnetically arrested disks is determined only by the presence or absence 
of strong vertical magnetic fields. The thermal and emission properties of 
plasma in these disks are not relevant in this respect.
Therefore, magnetically arrested disks can exist in a wide range of 
luminosities and accretion rates,
from significantly sub- to significantly super-Eddington rates.
This is unlike the case of optically thin ADAF models 
that become thermally unstable at moderate and high luminosities, 
$L\ga 10^{-3} L_{\rm Edd}$ (Abramowicz {\it et al.} 1995).

The magnetically arrested disk model can be used to explain low-frequency
($\sim 0.1-30$ Hz) 
QPOs that are often found in BH binaries in
the low/hard state (RM06).
The spiral accretion streams in magnetically arrested disks form patterns that 
have almost solid-body rotations with frequencies determined as a fraction 
($\sim 0.5$) of the Keplerian frequency at $R_{\rm m}$ (I08).
Relating this rotation with low-frequency QPOs, one can
estimate the QPO frequency,
\begin{equation}
\nu\sim 10^4\left({M\over M_\odot}\right)^{-1}
\left({R_{\rm m}\over R_g}\right)^{-3/2}\,{\rm Hz}.
\end{equation}
Note that Tagger \& Pellat (1999) and Tagger {\it et al.} (2004)
proposed a similar mechanism of the solid-body rotating structure
for low-frequency QPOs.
However, the nature of their structure, which is formed by
a spiral-wave generated because of the ``accretion-injection" instability,
is different from the nature of magnetically arrested disks.
Estimating $R_{\rm m}$ from eq.~(2) in the case of 
Cyg X-1, which is characterized by
$M\simeq 10\,M_\odot$ and $\nu\sim 1-3$ Hz (Rutledge {\it et al.} 1999),
one obtains $R_{\rm m}\sim 30\, R_g$.
The latter estimate agrees with the estimates of $R_{\rm tr}$ in Cyg X-1
obtained from interpretations of X-ray spectra
(Gierli\'nski {\it et al.} 1997; Axelsson {\it et al.} 2008).


\section{Transition Mechanism}

The flow that feeds an accretion disk orbiting the primary (BH) 
is supplied by the secondary (star) and, therefore, can carry a magnetic field,
whose strength and topology is determined by
the strength and topology of the stellar
magnetic field and relative orbital motions of binary components.
In this circumstance, the poloidal component
of the supplied fields can be inverted in time randomly
(depending on the secondary's magnetic activity) 
and/or quasi-periodically (with the time scale $\sim$ the orbital time).
The field inversion can result in a temporal disappearance of a magnetically arrested disk,
because of the annihilation of opposed fields, and
in a restoration of an optically thick disk that extends down
to the last stable circular orbit of the BH.
This can explain 
the transition between the low/hard state, 
which is characterized by the presence of a magnetically arrested disk, and the 
high/soft state, which is the manifestation of
an untruncated Shakura--Sunyaev--type disk.

The minimum magnetic flux that is required to form a magnetically arrested disk 
can be estimated
using equation~(1): 
\begin{equation}
\Phi_0\sim\pi R_g^2B_0\sim\pi R_g^2
\left({\dot{M}c\over\theta R_g^2}\right)^{1/2}
\approx 4\cdot 10^{20}\theta^{-1/2}\left({M\over M_\odot}\right)^{3/2}
\left({\dot{M}\over \dot{M}_{\rm Edd}}\right)^{1/2}{\rm Mw},
\end{equation}
where 
$\theta=H/R_g<1$, $H$ is the disk thickness near the BH horizon, and
$\dot{M}_{\rm Edd}=L_{\rm Edd}/c^2$.
For comparison, the magnetic flux in the Crab pulsar 
is estimated $\sim 10^{25} {\rm Mw}$.
The flux (3) comes from the outer
accretion radius $R_{\rm a}=2GM/v_{\rm a}^2$, 
where it is collected during the time $\tau$, so that
\begin{equation}
\Phi_0\sim B_{\rm a} R_{\rm a} v_{\rm a} \tau,
\end{equation}
where $B_{\rm a}$ and $v_{\rm a}$ are the magnetic induction and accretion
velocity at $R_{\rm a}$, respectively.
The estimate of $B_{\rm a}$, depending on the accumulation time $\tau$
and other parameters of the problem, can be obtained by substituting 
equation~(3) into (4), yielding
\begin{equation}
B_{\rm a}\sim 10^{-3} \theta^{-1/2} \left({v_{\rm a}\over c}\right)
\left({\tau\over {\rm year}}\right)^{-1}
\left({M\over M_\odot}\right)^{1/2}
\left({\dot{M}\over \dot{M}_{\rm Edd}}\right)^{1/2} {\rm G}.
\end{equation}
Let us consider the example of Cyg X-1 and use the following parameters:
$v_{\rm a}\sim 10^8$ cm/s,
$\tau\sim 1$ year, $\theta\sim 0.1$, and 
$\dot{M}\sim 0.1 \dot{M}_{\rm Edd}$.
Then, the estimate of the inner field yields $B_0\sim 10^8 {\rm G}$.
This field magnitude agrees with the magnitude
derived from polarimetric
observations of Cyg X-1 (Gnedin {\it et al.} 2003). 
The estimate of the outer field yields $B_{\rm a}\sim 10^{-5} {\rm G}$.
This estimate demonstrates how small the field supplied into the
accretion disk 
can be to initiate the formation of a magnetically arrested disk in Cyg X-1.
For comparison, the typical observed stellar fields are $\sim 1-100 {\rm G}$.

An evolution of the inverted fields in accretion disks has been studied
using axisymmetric 2D MHD simulations. 
Although these simulations do not correctly reproduce the 3D structure
of magnetically arrested disks (I08), they qualitatively correctly model 
the global evolution of the fields
and, therefore, are adequate here.
The employed numerical method is described in I08
and the simulation setup is assumed to be the same as in 
Model~B from there, except that the present simulations consider 
an inversion of the injected field. 
Figure~1 presents the evolution of magnetic fluxes in the midplane
of the model inside the five specific radii:
$210\,R_g$ ($=R_{\rm inj}$), $100\,R_g$, $50\,R_g$, $25\,R_g$, 
and $2\,R_g$ ($=R_{\rm in}$) (the black, red, green, blue,
and magenta curves, respectively).
The time is given
in units of the orbital time at $R_{\rm inj}$.
The spikes seen in the magenta curve are
due to a cycle accretion in the magnetically arrested region.
This cycle accretion is an artifact of the assumed axisymmetry
and is not present in 3D simulations (I08).
Initially, the magnetic fluxes are gradually increased with time 
because of the inward advection of the vertical field.
The moment of the field inversion is 
chosen at $t=5.1$, which is clearly distinguished 
as the abrupt jump of the black curve in Fig.~1.
Other curves, which correspond to the fluxes at smaller radii, 
are changed in time
with delays and more gradually.
The field inversion results in a temporal suppression of the accretion flow
at large radii (because more injected mass leaves the computational domain 
through the outer boundary) and development of a wide radial gap,
$\sim R_{\rm out}/2$,
between the outer edge of the ``old" disk 
and the inner edge of a ``new" disk, which carries the inverted field. 
As the mass accumulated in the old disk is reduced, the accretion rate
into the BH is also reduced. 
The increased time intervals between the spikes 
seen at $t\approx 6$ to $7.5$ in Fig.~1 
correspond to this reduction.
At $t\approx 7$, the new disk is fully developed and quickly fills
the radial gap, forming a continuous disk.
This disk has the vertical fields, which are inverted
at some disk radius.
Figure~2 represents this stage of evolution, showing the density
distribution and magnetic lines in the meridional cross-section
of the model at $t=8.52$. 
The central magnetic bundle is large and the flow
is magnetically arrested.
The inverted field (see Fig.~2b) closely approaches
the bundle, resulting 
in an intensive reconnection dissipation, which heats plasma
in a narrow interface between the opposed fields.
At $t\approx 9.5$, the central bundle is completely 
annihilated
and the dense disk extends all the way inward to $R_{\rm in}$.
The simulations show that
the accretion into the BH is continuous,
without signs of the cycle accretion, from $t\approx 9$ to 10.
Figure~3 illustrates this evolution stage, showing the model at $t=10.08$.
The simulations were finished at $t=10.75$. At this moment,
the inverted field builds up another central magnetic 
bundle, which arrests the accretion flow. 
It becomes apparent that the further evolution of this model 
will basically repeat the initial evolution.

The SPL state is similar in some respects to the high/soft state (see Sec.~1).
This motivates us to explain the SPL state as a final phase of the more
extended period of field annihilation that precedes the high/soft state.
Unfortunately, no supporting 3D simulations of this period have been done yet
and our discussion is only limited by the following qualitative statements.
(1) At the final phase, the radius $R_{\rm m}$ is small
and the optically thick disk can extend deep inside, providing
a sizable spectral thermal component.
(2) A significant fraction of the disk emission can be provided by the reconnection
dissipation of the stored magnetic energy. This can explain the specific
steep power law spectral feature.
(3) Small $R_{\rm m}$ means that dense blobs and accretion streams,
similar to that found by I08, can be developed in the vicinity of the BH horizon.
Orbital motions of such blobs and streams near the last stable circular orbit of the BH
can explain high-frequency ($\sim 150-450$ Hz) QPOs
(e.g., Stella \& Vietri 1999; Abramowicz \& Klu\'zniak 2003).

\section{Summary}

This letter proposes a detailed mechanism for emission state transition
in BH X-ray binaries.
The mechanism is based on a dynamical model of a magnetically arrested disk
obtained in recent 3D MHD simulations (I08) that can describe 
the low/hard state.
In this state, the central region of an optically thick
accretion disk orbiting
the BH is filled with a strong (equipartition)
vertical magnetic field that arrests the flow,
resulting in a truncated disk.
The accretion, however, is not suppressed in this disk
but takes the form of thin spiral streams that penetrate through
a highly magnetized, hot, low-density plasma.
The spiral flow twists the vertical field and produces powerful Poynting jets.
Radiation from the disk is dominated by hard X-rays
emitted by the hot plasma and
includes a radio component from the jets.
The magnetically arrested disk is capable to explain low-frequency QPOs.

The transition from the low/hard to high/soft state occurs
when the magnetic field at the outer boundary is inverted.
An annihilation of this field, carried inward by the flow,
and the central field results
in a temporal disappearance of the magnetically arrested disk.
For a moment, the optically thick disk extends inward to
the last stable orbit of the BH and no jets are produced.
This represents the high/soft state, which ends
as soon as the continuous supply of
the inverted field
results in the formation of another magnetically arrested disk.
The SPL state and associated with it high-frequency QPOs develop in
this scenario at the period of intensive annihilation of the inverted fields,
which precedes the high/soft state.
Quasi-periodic or random inversion of magnetic fields
in the accretion flow supplied by the secondary star
can explain the observed spectral
transitions of BH X-ray binaries.



\clearpage

\begin{figure}
\epsscale{1.0}
\caption{Evolution of vertical magnetic fluxes (in arbitrary units)
in axisymmetric 2D simulations of the accretion disk model.
The five curves shown correspond to the fluxes through the equatorial plane
inside five fixed radii: $210 R_g$ ($=R_{\rm inj}$, black),
$100 R_g$ (red), $50 R_g$ (green), $25 R_g$ (blue), and
$2 R_g$ (magenta).
The time is in units of the orbital time at $R_{\rm inj}$.
The injected field is inverted at $t=5.1$.
\label{fig1}}
\end{figure}

                                                                                
\begin{figure}
\epsscale{1.0}
\caption{Distribution of density (a) and magnetic lines (b) 
in the meridional plane at $t=8.52$
from axisymmetric 2D simulations of the accretion disk model.
The black hole is located on the left side of the images, 
and the small open circles there
correspond to the inner boundary around the black hole at $R_{\rm in}=2R_g$.
The axis of rotation is in the vertical direction.
The domain shown has the radial extend $R_{\rm out}=220 R_g$
along the equatorial plane and vertical extend from
$-R_{\rm out}/2$ to $R_{\rm out}/2$.
The color bar on the right in (a) indicates the scale for log$\,\rho$
(in arbitrary units). The lines in (b) have been plotted using the
method of Cabral \& Leedom (1993).
The field carried inward by the disk 
and the field in the central magnetic bundle have inverted 
vertical components (b).
The central bundle arrests the accretion flow and 
forms a truncated disk (a).
\label{fig2}}
\end{figure}


\begin{figure}
\epsscale{1.0}
\caption{Same as in Figure~2, but at $t=10.08$.
The central magnetic bundle seen in Figure~2b has disappeared because of the
annihilation with the incoming inverted field (b). The disk is untruncated
and extends all the way inward to $R_{\rm in}$ (a).
\label{fig3}}
\end{figure}

\clearpage

\end{document}